# Fogbanks: Future Dynamic Vehicular Fog Banks for Processing, Sensing and Storage in 6G

A.A. Alahmadi[1], M.O.I. Musa[1], T.E.H. El-Gorashi[1], J.M.H. Elmirghani[1], S. Grant-Muller[2], D. Hutchison[3], A. Mauthe[3], M. Dianati[4], C. Maple[4], L. Lefevre[5] and A. Lason[6]
[1]University of Leeds (EEE), UK; [2]University of Leeds (Institute for Transport Studies), UK; [3]Lancaster University, UK; [4]University of Warwick, UK; [5]Inria, France; [6]AGH, Poland

**Abstract**

*Fixed edge processing has become a key feature of 5G networks, while playing a key role in reducing latency, improving energy efficiency and introducing flexible compute resource utilization on-demand with added cost savings. Autonomous vehicles are expected to possess significantly more on-board processing capabilities and with improved connectivity. Vehicles continue to be used for a fraction of the day, and as such there is a potential to increase processing capacity by utilizing these resources while vehicles are in short-term and long-term car parks, in roads and at road intersections. Such car parks and road segments can be transformed, through 6G networks, into vehicular fog clusters, or "Fogbanks", that can provide processing, storage and sensing capabilities, making use of underutilized vehicular resources. We introduce the Fogbanks concept, outline current research efforts underway in vehicular clouds, and suggest promising directions for 6G in a world where autonomous driving will become commonplace. Moreover, we study the processing allocation problem in cloud-based Fogbank architecture. We solve this problem using Mixed Integer Programming (MILP) to minimize the total power consumption of the proposed architecture, taking into account two allocation strategies, single allocation of tasks and distributed allocation. Finally, we describe additional future directions needed to establish reliability, security, virtualisation, energy efficiency, business models and standardization.*

## Introduction

Cloud computing has redefined the computation and communication environment by utilizing multiple resources such as servers, storage devices, and other network hardware to provide on-demand services to end users with high reliability and scalability at a lower cost. The huge growth in increasingly resource hungry cloud-based applications calls for more research into new architectures and solutions to offload the computational burden in the centralized data centers, and to improve the performance of applications. Conventional cloud data-centers are usually centralized and are accessed through the Internet [1]. This centralized cloud structure faces several challenges such as single point failure, reachability, and transmission latency. An alternative distributed cloud can be used in which a group of resources located at the edge of the network (either stationary or mobile) provides the same concept of on-demand services to the end user [2]. Distributed cloud platforms can be composed from any available user-owned resources that allow processing, storage, networking and sensing. Following this concept, vehicular clouds can be formed if the vehicle on-board processing, storage, and sensing devices are clustered together to form short-term fog units composed of many vehicles (each vehicle effectively acting as a server in a mobile micro data centre) at the edge of the network [3]. Furthermore, when mobile, the sensing capabilities of vehicles can be used as a form of mobile IoT platform with sensors that may include cameras, pollution sensors, traffic flow sensors and road surface sensors among others.

In this article we introduce Fogbanks, a vision that can help transform edge processing, storage/caching and sensing in 6G communication networks through the use of the capabilities of distributed vehicles to form intelligent vehicular fog processing clusters at the edge of the network. These capabilities are expected to grow significantly with the introduction of autonomous vehicles in the near future. Currently, vehicles continue to be used typically for 2-4 hours per day [3]. Therefore, a significant portion of the processing capabilities of such autonomous and connected vehicles remain unused during the day. Cars in an enterprise car park may number in the hundreds to thousands and remain in the enterprise car park for typically 7-8 hours per day. If vehicles are connected in such a car park, (using wireless or a fibre cable integrated with the charging cable and its plug) their processors (typically 2-10 processors per vehicle) can be networked, thus transforming the car park into a significant edge processing micro data centre. A set of Fogbanks made up of the parking rows and floors in a car park can thus be composed as Figure 1 shows. Similarly, cars in airports may be parked for one to two weeks, making the capabilities of such vehicles available to transform such car parks to processing units / fog units at the edge of the network on a semi-permanent basis as departing cars are replaced. Fixed fog processing can be used to augment the facility and hence help provide minimum reliability and availability guarantees. Additionally, cars are likely to be parked in city car parks for 2-4 hours typically, providing small clusters of edge processing nodes. On shorter time scales, vehicles may spend an hour or less at charging stations, thus providing opportunities to form Fogbanks at charging stations. On the very shortest time scales, clusters of vehicles may be formed at traffic intersection points where the traffic light may own a computational problem and may assign chunks of such a computational problem to vehicle clusters at the intersection. The clusters report results before departing the intersection. At busy intersections in cities, typically at least one traffic stream is stationary, thus providing opportunities to distribute computational tasks to nearby processors. These vehicles thus have the potential to form efficient distributed computational resources at the edge of the network, much closer to the source of the problem.

The networks and computational resources are highly dynamic, however, with time constants that can range from minutes to weeks. Therefore, appropriate network architectures, network algorithms and transport studies are needed to understand these new forms of dynamic distributed computational resources. It is also essential to design key features in such networks including energy efficiency, low latency, high reliability and high availability. The envisaged new form of Fogbanks edge processors can thus reduce the cost of providing the computational services needed by making use of underutilized resources in vehicles and can enable new services.

## Related Work on Vehicular Fog

Research efforts to optimize vehicular clouds have focused, to date, on four main aspects; namely resource management, service provision, dependability and security.

Managing vehicular resources is challenging due to the inherent heterogeneity of resources, the mobility of vehicles in dynamic scenarios, and the random arrival/departure times of vehicles in static scenarios. The authors in [4] proposed a stochastic model to predict the availability of computational resources of parked vehicles by considering a Poisson distribution for vehicle arrivals and departures. They presented a mathematical reasoning analysis of the probability distribution of a long-term parking lot occupancy as a function of time and conducted a simulation experiment to validate their analytical results.

Vehicular computing resources can be grouped into clusters to collectively serve demands. Cluster formation and cluster head selection techniques can improve the service provisioning capabilities of vehicular clouds. A distributed dynamic clustering algorithm was presented in [5] to group vehicles into clusters in a vehicular cloud-based data gathering and delivery service. The sizes of these clusters vary with the mobility of vehicles to minimize delay and maximize throughput. Virtualization is an essential enabler of resource management in the inherently heterogeneous vehicular clouds. The allocation of virtualized computational resources of vehicular clouds is optimized in [6] to minimize the cost and maximize the income of vehicular cloud services in the long term.

In [7] the task assignment problem in vehicular clouds is formulated as an optimization problem, taking the heterogeneity of vehicular resources and the interdependency of computing tasks into consideration and a modified genetic algorithm is proposed to solve the scheduling problem. In [8] a mixed integer linear programing (MILP) formulation of the task assignment problem is presented considering migration to ensure service continuity under dynamic resource availability of vehicular clouds. The dynamicity of vehicular clouds is also addressed in [9], with the solution employing reinforcement learning in vehicular clouds to adapt the resource management techniques so as to select a vehicle to serve the demand given the variation in the vehicular cloud.

Energy efficient resource management of a distributed cloud system is investigated in [10] by offloading some of the central cloud load to vehicular clouds. A MILP model is formulated to minimize the total power consumption (network and processing) by optimizing the assignment of tasks to the central cloud and/or the vehicular cloud, taking into account the impact of task splitting among different clouds.

Vehicular clouds storage and computational resources can be leveraged jointly with the central cloud, or core cloud to provide different services. A vehicular cloud on-demand service provisioning framework is developed in [11]. The authors proposed three different game theory-based approaches to minimize the service cost and latency and maximize privacy. In [3] data dissemination in a vehicular cloud of limited storage and intermittent connectivity is supplemented by prefetching data from the cloud to roadside access points; also an online learning algorithm is proposed to optimize the data prefetching.

The volatility of vehicular cloud resources creates concerns about service reliability. The authors in [12] considered redundant assignment strategies to serve demands in a vehicular cloud. They presented a comprehensive study of the mean time to failure of these strategies through theoretical analysis and simulations. A reliable serving vehicle selection algorithm was proposed in [13]

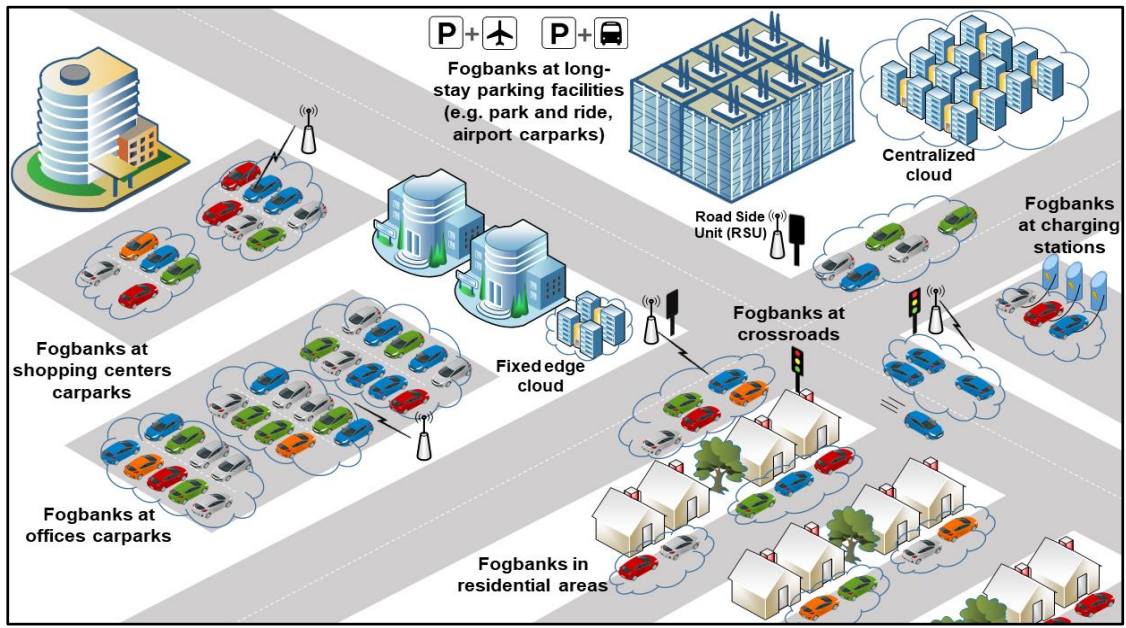

*Figure 1* Fogbanks: Dynamic vehicular fog networks for processing, sensing and storage, with fogbanks formed as shown in long stay multi-level car parks, shopping centres car parks, and at charging stations and road intersections

where roadside units support vehicular resources to guarantee task completion in the case of vehicle departure. Security challenges in vehicular clouds were identified and analysed in [14]. A security framework that tackles authentication under mobility and scalable security for vehicular clouds was also presented. Table 1 gives a summary of the above research efforts in vehicular clouds.

| Topic | Reference | Key Contribution |
|---|---|---|
| Resource Management | [4] | Proposed a stochastic model to predict the availability of resources in parked vehicles scenarios. |
| | [5] | Proposed a distributed dynamic vehicle clustering algorithm. |
| | [6] | Optimized the allocation of virtualized vehicular resources to minimize the cost and maximize the income. |
| | [7] | Optimized task assignment considering the heterogeneity of vehicular resources and tasks interdependency. |
| | [8] | Investigated the use of task migration to ensure service continuity. |
| | [9] | Employed reinforcement learning to adapt the vehicle selection techniques to variation in the vehicular cloud. |
| | [10] | Studied energy efficient task assignment in a distributed cloud scenario by offloading some of the central cloud load to vehicular clouds. |
| Service provisioning | [11] | Considered game theory based approaches to minimize the service cost and latency and maximize privacy. |
| | [3] | Used data prefetching from the cloud to roadside access points in data dissemination services in a vehicular cloud of limited storage and intermittent connectivity. |
| Reliability | [12] | Considered redundant assignment strategies to improve reliability |
| | [13] | Proposed a reliable vehicle selection algorithm. |
| Security | [14] | Presented a security framework to tackle authentication under mobility and scalable security for vehicular clouds. |

*Table 1* Summary of research efforts in vehicular clouds resource management, service provisioning, reliability and security

**Processing allocation in cloud-supported Fogbanks Architecture**

Figure 2 shows a potential Fogbank end-to-end architecture, where the short-term vehicular nodes (VNs) are clustered as a vehicular Fogbank (VF). These Fogbanks can exist as a result of any of the vehicular cluster schemes presented in Figure 1. The architecture is supported by fixed fog nodes located at the access layer and connected to the optical line terminal (OLT) and optical network unit (ONU). These fixed fog nodes are referred to as OLT fog (LF) and ONU fog (NF) nodes. The existence of the fog nodes at the access layer guarantees a minimum level of service in the absence of the vehicular nodes (Fogbank), or when the Fogbanks are thin. Furthermore, the presence of the LF and NF ensures low latency is supported together with energy efficiency. Beside the Fogbank and the fixed fog nodes, the architecture is supported by a central cloud (CC) to provide the conventional support needed for applications where latency is not a significant constraint and to absorb demands beyond those that can be supported at the edge of the network.

Tasks originating from nearby devices are collected at the edge layer by the roadside units (RSU). The RSU allocates these tasks to the available processing nodes (CC, LF, NF, and VF) based on the executed allocation strategy. The processing allocation in this study is optimized using mixed integer linear programming (MILP) to minimize the overall power consumption of the end-to-end architecture. Other optimization metrics can also be considered such as the minimization of latency, and minimization of cost and combinations of these metrics with power consumption. The formulation can follow a similar approach, with the background and detailed MILP formulations discussed in [1], [3]. Here two allocation strategies are considered where tasks are split over many processing nodes (distributed allocation) or allocated to only one processing node (single allocation). Consideration is also given here to different scenarios based on the demand size, vehicle nodes densities and Fogbank size.

The objective of this MILP model is to minimize the total power consumption which is comprised of the processing and networking power consumption, as given below in equation (1).

Objective: Minimize
$$\sum_{p \, \epsilon \, PN} \sum_{k \, \epsilon \, K} X_{kp} \, E_p \; + \sum_{n \, \epsilon \, N} \sum_{k \, \epsilon \, K} F_k \, \Psi_n \quad (1)$$

The first part of the equation calculates the processing power consumption of each processing node $p$ allocated task $k$, where $X_{kp}$ is the workload demanded by task $k$, in million instructions per second (MIPS), assigned to processing node $n$, $E_p$ is the power in watts per MIPS of the processing node, calculated using the maximum processing capacity of the node. The second part of the equation calculates the power consumption that results from sending the traffic. This power consumption results from the workload demand of task $k$ as it passes through each networking device $n$. The traffic demand of task $k$, is denoted by $F_k$ while $\Psi_n$ denotes the power per Mbps of the networking device $n$, also calculated using the maximum networking capacity of the node.

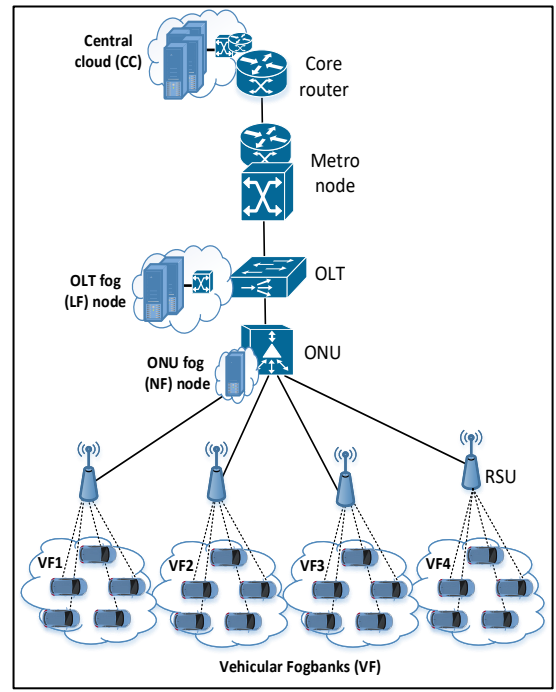

*Figure 2* Cloud supported Fogbanks architecture

This optimization problem is solved while observing the following constraints:
1) The summation of processing workloads allocated to a node must falls within the node processing capacity.
2) The traffic associated with the task must be less than the link capacity leading to the allocated processing node.
3) Each task is allocated to one processing node (in case of single allocation).

Table 2 summarise the optimisation model input parameters for the tasks and processing nodes.

| parameter | Value |
| --- | --- |
| Number of generated tasks | 50 |
| Processing task demand in MIPS | 500 – 5000 MIPS |
| Task required data rate | 5 – 50 Mbps |
| Number of central clouds (CC) | 1 (incapacitated) |
| Number of OLT fog nodes (LF) | 1 (54400 MIPS capacity) |
| Number of ONU fog nodes (NF) | 1 (6000 MIPS capacity) |
| Number of vehicular Fogbanks (VF) | 4 VFs |
| Fogbank capacity (vehicles density) | 5 – 15 vehicles per VF (3200MIPS each) |

**Table 2** *Input parameters for the tasks and processing nodes.*

The following considerations describe the scenario evaluated through MILP:
1) All VNs located within the same Fogbank can communicate with one associated RSU and cannot communicate with other RSUs directly.
2) All tasks are generated from nearby devices located within one vehicular Fogbank area ($VF_i$). Therefore, the RSU in $VF_i$ is considered the main controller to choose the optimal processing placement.
3) All RSUs communicate with each other through ONU and cannot communicate with each other directly.
4) The processor of CC has the best processing energy efficiency, followed by LF processor, NF processor, and finally vehicle processor. However, VNs have the minimum hop route and hence the best networking energy efficiency, followed by NF, LF, and finally CC.
5) In both allocation strategies, individual tasks can be allocated to different processors. However, in the distributed allocation setting, one task can be split among different processors based on the optimal decision. On the contrary, in the single allocation strategy, each task must be allocated in only one processing node.

For each allocation strategy (single and distributed), four test scenarios are considered. In the first scenario, (CC), all tasks are allocated to the central cloud. This scenario is considered as a baseline where fixed fog or Fogbanks do not exist. The second scenario considers a full architecture, including cloud and fog. However, it is assumed that no vehicular nodes are present. This also take into account some periods where vehicles are not present in the area (off-peak day time). The third scenario considers a low VNs density with the assumption that each Fogbank is comprised of five VNs, thus a total of 20 vehicles. The fourth scenario assumes a high VNs density where the total number of vehicles increases to 60 vehicles (15 VNs per VF). The last two scenarios are considered to test the impact of the VNs density on the processing allocation and therefore the total power consumption of the architecture.

Figures 3(a) and 3(b) show the total power consumption of the single and distributed allocation strategies, respectively. Figure 4 presents the processing allocation size in each processing node, which reflects the power consumption values in Figure 3. As presented in Figures 3(a) and 3(b), the total power consumption in CC allocation shows a gradual increase with increase in the processing demand. At 3500MIPS, the power shows an abrupt increase due to the fact that the amount of processing allocated at the cloud is fulfilled by more than one server. This case is not presented in Figure 4 as all tasks are allocated at one location (CC).

In the second scenario, where the Fogbanks capacity is zero, the power consumption has a better result at the low demands (500-1500MIPS) as tasks are assigned to LF. Although the NF, is more efficient in terms of total power consumption compared to LF, assigning all the tasks to one location saves more power due to the power overhead that results from activating servers at different locations. This causes high demand tasks (beyond 1500MIPS) to be assigned to one server at CC rather than activating the other two fog nodes (LF and NF). Despite this, in some cases where demands require more than one server in the

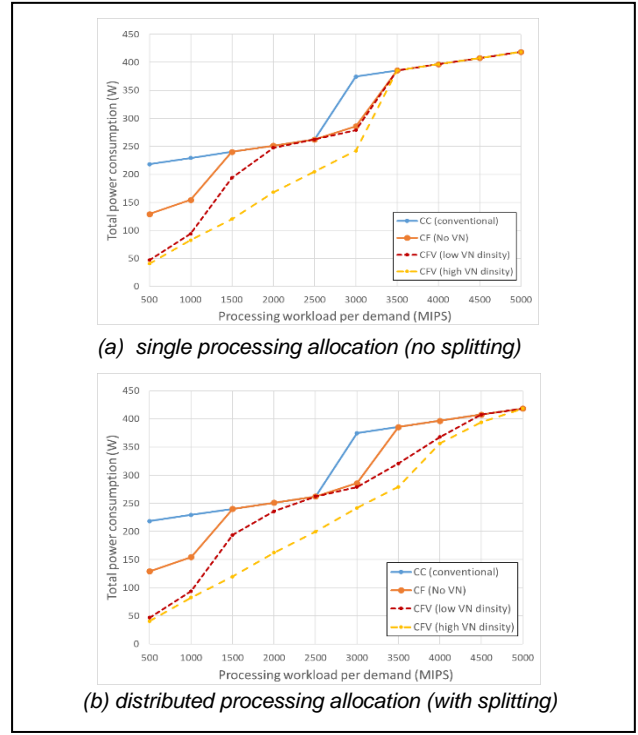

*(a) single processing allocation (no splitting)*

*(b) distributed processing allocation (with splitting)*

*Figure 3 Total power consumption, with increasing processing workload per demand, for four scenarios: (i) central cloud (CC), (ii) Cloud and fog (C/F) with no vehicular nodes, (iii) C/F with low vehicular nodes density, and (iv) C/F with high vehicular nodes density.*

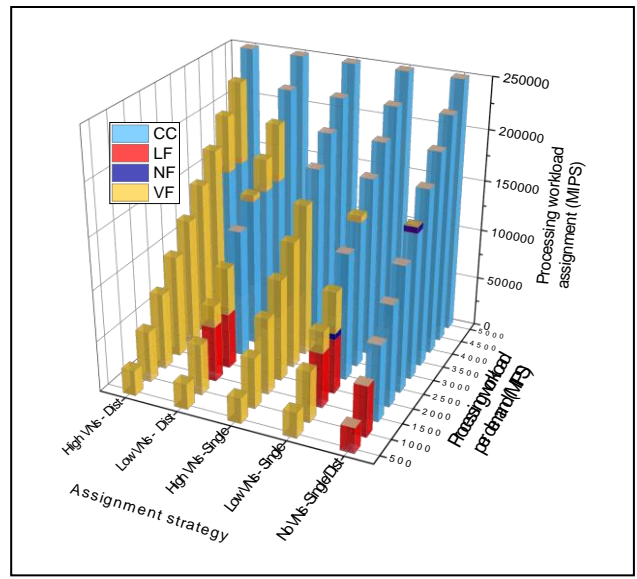

**Figure 4** *Processing workload allocated at each processing node in each scenario.*

CC, it is more efficient to activate the server at LF or NF and allocate the remaining tasks to the smallest sufficient processing node at the fog network (i.e. at 3000MIPS).

In the third scenario, where the number of vehicles increased to a total of 20 VNs, the power consumption decreased in both allocation strategies, single and distributed. In the single allocation case, Figure 3(a), the power consumption is reduced by 64% compared to the case where no VNs were available and is reduced by up to 79% compared to CC allocation. However, when the demands size increases beyond 3000MIPS, no tasks are allocated in the Fogbanks, as the VN processor (with 3200MIPS) becomes insufficient for the demands requirements. On the other hand, in Figure 3(b) where distributed allocation is applied, demands beyond the VNs size are split between different VFs or between VF and other fixed fog nodes. This results in lower power consumption, a maximum reduction in power consumption of 17% compared to the single allocation. However, with high workload demands, CC becomes more efficient due to its ability to accommodate all the tasks rather than distributing them among different processors in the lower fog network, including the vehicles. The reason for this was explained earlier as activating one CC processor to accommodate all tasks becomes more efficient than activating many smaller processors in the fog layer.

In the final scenario, where the VN density increased to a total of 60 vehicles, both allocation strategies exhibited enhanced energy efficiency which results as more tasks are allocated in the existing vehicles. As depicted in Figure 3(a), the single allocation strategy achieved between 12% and 38% power savings with increased VNs density, compared to the low VNs density. This saving is limited to the cases where demands are within the VNs capacity. This explains the increase in power consumption, at 3500MIPS, where all tasks are allocated to the CC, and therefore in this case the power consumption is the same as that of the conventional baseline results, regardless the existence of the fixed fog nodes and vehicular Fogbank. On the contrary, in the distributed allocation strategy in Figure 3(b), tasks are allocated to Fogbanks even beyond the 3500MIPS. This strategy is optimum when the capacity of one server in the CC is exceeded and the addition of more CC servers adds substantial power consumption. Here the Fogbanks become more efficient, whenever they are able to accommodate the remaining high demand tasks. All the allocation decision in each processing nodes are summarized in Figure 4.

Part of the aim of this work, is to design a dynamic architecture that is able to increase the Fogbank size and connect many Fogbanks together in the cloud-supported architecture. Accordingly, it is very important to study the allocation behaviours among the Fogbank clusters considered. Figure 5 summarizes the processing allocation for each individual vehicular Fogbank, taking into account that tasks were generated from one Fogbank cluster (VF1).

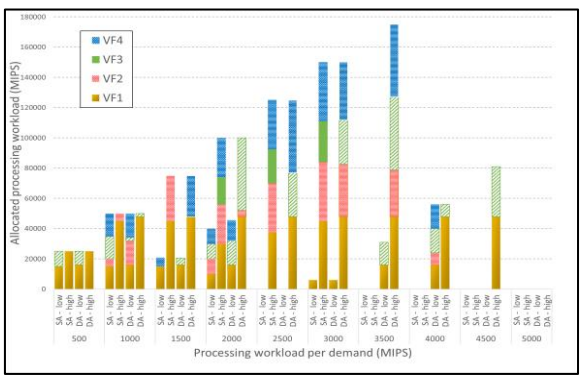

*Figure 5* processing workload allocated at each vehicular Fogbank (VF) in single allocation strategy with low and high VN densities (SA-low) and (SA-high), and distributed allocation strategy with low and high VN densities (DA-low) and (DA-high)

The results show that the majority of tasks that were allocated to the local Fogbanks, were tasks generated from the same cluster, as this allocation costs the least power consumption. For the remaining tasks that were not satisfied by the local VF, three options are available for the allocation decision:

1) If the NF capacity is enough to accommodate the remaining tasks, the model prioritizes this fog node over the other Fogbanks. This is because the RSU controller cannot communicate directly with other Fogbanks, but it can do over the ONU. This makes the NF more efficient in networking power consumption than the other Fogbanks.
2) If the NF cannot satisfy the remaining workload, and other available Fogbanks can, then these tasks are allocated to the available vehicles in the non-local Fogbanks.
3) Lastly, if the available Fogbanks cannot fully allocate the required workload then all tasks will be sent to the LF or CC.

Moreover, in the case where processing tasks are split between several locations (Distributed allocation), it is more efficient to split the tasks among vehicles located in the same Fogbank, rather than splitting the task between different Fogbanks. However, accommodating tasks into one location saves more power than allocating different tasks to different Fogbanks.

**Future Directions for Designing Fogbanks**

The Fogbanks concept holds significant promise for edge processing in 6G networks with the introduction of autonomous and connected vehicles which will introduce significant additional computational resources that can be exploited in the network. Several key issues have to be addressed in these future edge processing 6G networks:

**1. Transport studies:** This is a key feature lacking in current work. Current studies on vehicular clouds assume statistics for the arrival and departure of vehicles in certain road intersections and car parks. Detailed transport studies are needed alongside physical layer wireless and optical communications (e.g. fiber cable integrated in vehicle charging cable plugged at carpark or at charging station) and computing studies into Fog formation. These transport studies can model the flow of vehicles in cities using real city layouts and calibrated transport models used in vehicular flow studies and in city planning.

**2. Dynamic nature of Fogbanks:** Algorithms and mechanisms have to be developed to form and to manage the dynamic Fogbanks, based on accurate understanding of the vehicular flow and the transport layer models. With the availability of these large data sets, machine learning and artificial intelligence algorithms can be used to manage the dynamic Fogbanks.

**3. Awareness and perception:** The autonomous vehicles forming Fogbanks should possess awareness of the environment and perception including the presence of other vehicles in the neighborhood and their capabilities.

**4. Dependability:** This presents new challenges as the usual communications setting assumes that the network resources, including computational, sensing and storage resources are deterministic, known and available (most of the time, e.g. five nines). The availability of these resources is stochastic in nature in Fogbanks posing new challenges. The use of fixed Fog / fixed edge processing as in Figure 1 can help, with Fogbanks, for example providing additional resources.

**5. Virtualization and energy efficiency:** Processing, storage and sensing are growing in importance and as such Fogbanks can provide avenues for standard operators and Mobile virtual network operators (MVNOs) to provide services beyond communications, by slicing and providing edge resources based on the joint capabilities of Fogbanks and fixed edge clouds. The proximity of Fogbanks to end-users can provide additional energy efficiency benefits, where it becomes economical for example to use a rural car park to provide edge processing.

**6. Incentives and business models, market penetration:** Uptake will likely rely on incentive mechanisms, business models and market penetration. For example, operators may establish financial incentives that may be awarded to autonomous vehicle owners who participate by making some of their spare vehicle processing capabilities available. Rewards can be established with city councils and municipalities to reduce city taxes or offer free parking to participating vehicles.

**7. Standardization:** Standardization is key for the uptake of the Fogbanks concept in terms of network architecture, routing protocols and algorithms for the establishment, management and disintegration of Fogbanks. In addition, standardization of usage, measurement and monitoring are essential for rewards and reliability.

**8. Additional areas:** The Fogbanks concept opens many additional avenues for research and exploitation that are hard to cover exhaustively, however applications and implications on the physical layer (and opportunities, e.g. mobile caching and mobile processors) are also worth further attention.

## Conclusions

We have introduced the idea of Fogbanks where the significant spare processing, storage, networking and sensing capabilities seen now in connected vehicles and expected to increase in autonomous vehicles can be exploited. We use these resources to form edge processing banks, Fogbanks, when vehicles are in carparks, at road intersection or at electric vehicle charging stations for example. Unlike the initial approaches, reviewed, where vehicular clouds are formed at low density, in Fogbanks, it is expected that the significant processing capabilities in autonomous vehicles and the increase in the number of connected vehicles will lead to thick Fog (Fogbanks) at the edge of the network that can be used for processing and sensing for example. We studied the processing allocation problem in a cloud assisted Fogbanks architecture. We solved this allocation problem using MILP optimization to minimize the total power consumption of the proposed architecture, taking into account two allocation strategies and different VNs densities. The results showed a power saving by 80% when Fogbank clusters are introduced compared to the central cloud processing. Moreover, increasing the VNs density, saves 38% to 64% power compared to the cases where only fixed fog nodes are used at low VNs density. The ability to split a task over different vehicles and Fogbank clusters helps to utilize the participant vehicles processing resources efficiently, thus reducing power consumption by up to 28%, compared to the single allocation strategy. Finally, several key issues and future directions are discussed for Fogbanks edge processing in 6G networks.

## Acknowledgements

We would like to acknowledge funding from the Engineering and Physical Sciences Research Council (EPSRC) for the INTERNET (EP/H040536/1), STAR (EP/K016873/1) and TOWS (EP/S016570/1) projects. All data are provided in full in the results section of this paper.